\documentclass[%
 reprint, 
 amsmath,amssymb,
 aps,
prl,
floatfix,
]{revtex4-1}
\usepackage[normalem]{ulem}
\usepackage{amssymb}
\usepackage{graphicx}
\usepackage{epsfig} 
\usepackage{dcolumn}
\usepackage{bm}
\usepackage{epstopdf}
\usepackage{multirow}
\usepackage{amsmath}
\usepackage{booktabs}
\usepackage{color}
\usepackage{gensymb}
\usepackage{dcolumn}
\usepackage{kantlipsum}
\usepackage{hyperref}
\usepackage{ulem}
\usepackage{braket}
\usepackage{physics}
\usepackage{comment}
\usepackage[utf8]{inputenc}
\usepackage[T1]{fontenc}
\usepackage{mathptmx}
\usepackage{natbib}
\usepackage{float}
\usepackage{mathtools}

\newcolumntype{.}{D{.}{.}{-1}}
\begin{document}

\title{Dressed-State Spectroscopy and Magic Trapping of Microwave-Shielded NaCs Molecules}

\preprint{APS/123-QED}

\author{Siwei Zhang}
\affiliation{Department of Physics, Columbia University, New York, New York 10027, USA}
\author{Weijun Yuan}
\affiliation{Department of Physics, Columbia University, New York, New York 10027, USA}
\author{Niccol\`{o} Bigagli}
\affiliation{Department of Physics, Columbia University, New York, New York 10027, USA}
\author{Claire Warner}
\affiliation{Department of Physics, Columbia University, New York, New York 10027, USA}
\author{Ian Stevenson}
\affiliation{Department of Physics, Columbia University, New York, New York 10027, USA}
\author{Sebastian Will}\email{Corresponding author. E-mail: sebastian.will@columbia.edu}
\affiliation{Department of Physics, Columbia University, New York, New York 10027, USA}
\date{\today}
\begin{abstract}
We report on the optical polarizability of microwave-shielded ultracold NaCs molecules in an optical dipole trap. While dressing a pair of rotational states with a microwave field, we observe a marked dependence of the optical polarizability on the intensity and detuning of the dressing field. To precisely characterize differential energy shifts between dressed rotational states, we establish dressed-state spectroscopy. For strong dressing fields, we find that a magic rotational transition can be engineered and demonstrate its insensitivity to laser intensity fluctuations. The results of this work have direct relevance for evaporative cooling and the recent demonstration of molecular Bose-Einstein condensates [Bigagli, \textit{et al.}, Nature (2024)] and may open a door to precision microwave spectroscopy in interacting many-body systems of microwave-shielded molecules.
\end{abstract}

\maketitle

Collisional shielding via microwave fields~\cite{gorshkov2008suppression, cooper2009stable, karman2018microwave,lassabliere2018controlling, anderegg2021observation,schindewolf2022evaporation, bigagli2023collisionally, lin2023microwave} has led to a recent surge of results in the field of ultracold molecules.
Microwave shielding relies on microwave dressing~\cite{yan2020resonant} to engineer a repulsive collisional barrier between molecules. 
The suppression of inelastic losses under microwave shielding has led to the realization of a degenerate Fermi gas~\cite{schindewolf2022evaporation} and the first Bose-Einstein condensate (BEC) of dipolar molecules~\cite{bigagli2024observation}. Furthermore, microwave shielding provides a powerful tuning knob to control intermolecular interactions~\cite{micheli2007cold, gorshkov2008suppression, cooper2009stable, lassabliere2018controlling, karman2024upcoming}, as demonstrated in several recent experiments \cite{yan2020resonant, schindewolf2022evaporation, chen2023field, bigagli2024observation}, and is poised to become a workhorse technique for the preparation, control, and manipulation of many-body systems of ultracold molecules. 

While previous work on microwave dressing has focused on its impact on the collisional properties of molecules, the coupling of internal rotational states via the dressing field is also expected to show effects on the single molecule level. In particular, the microwave-induced mixing of rotational states should modify the optical polarizability of dipolar molecules. On the other hand, the control of optical polarizability of molecules has long been a topic of interest. In order to realize long rotational coherence times in optically trapped molecules, approaches have been developed that minimize the differential polarizability between rotational states. These include the tuning of the optical trapping field to a magic polarization angle \cite{neyenhuis2012anisotropic, blackmore2020controlling, burchesky2021rotational, lin2021anisotropic}, a magic polarization ellipticity \cite{park2023extended}, or a magic wavelength \cite{bause2020tune,gregory2024second}. Through the controlled manipulation of rotational quantum states, microwave dressing opens a new possibility in controlling single molecule properties like differential optical polarizability.

In this letter, we investigate the interplay between microwave dressing and optical polarizability with ultracold NaCs ground state molecules and discuss novel opportunities arising from it. First, we observe a variation of trap depth and trap frequency on $\Delta / \Omega$, where $\Delta$ is the detuning and $\Omega$ is the Rabi coupling of the microwave dressing field. Second, we develop dressed-state spectroscopy, precisely measuring energy shifts and demonstrating coherent Rabi oscillations between dressed rotational states. Finally, utilizing dressed-state spectroscopy, we show that `magic' transitions can be engineered via strong microwave dressing fields that do not exist in the absence of dressing. 

\begin{figure}[]
    \centering
    \includegraphics[width = 8.6 cm]{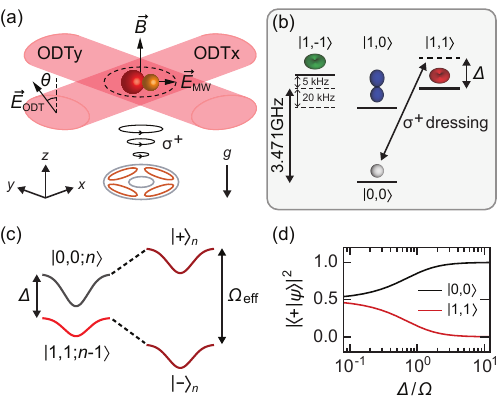}
    \caption{Microwave-shielded molecules and optical trapping. (a) Illustration of the experimental setup with microwave-shielded NaCs molecules trapped in a crossed optical dipole trap. Dipole trap beams ODTx and ODTy propagate along the $x$- and $y$-direction, respectively. The beams differ in laser frequency by 160 MHz.  $\theta$ denotes the angle between the optical electric field $\vec{E}_{\mathrm{ODT}}$ and the vertical magnetic field $\vec{B}$. (b) Rotational structure of NaCs ground state molecules at $B = 864$ G. A blue-detuned $\mathrm{\sigma^+}$-polarized dressing field couples states $\left | 0, 0 \right>$ and $ \left | 1,1 \right>$. (c) Modification of the optical trapping potential under microwave dressing. Microwave coupling leads to dressed states $|+\rangle$ and $|-\rangle$ with an energy splitting $\Omega_{\mathrm{eff}} = \sqrt{\Omega^2+\Delta^2}$ and a modified optical polarizability. $\left | J,m_J;n \right>$ denotes the product state of $\left | J,m_J \right>$ and the Fock state $\left | n \right>$, where $n$ denotes the number of photons in the microwave field and $J$ is the rotational angular momentum and $m_J$ its projection on the $z$-axis. (d) State composition of $\left | + \right>$ as a function of microwave shielding parameter $\Delta/\Omega$.}
    \label{fig:1}
\end{figure}

We start with a theory discussion on the optical polarizability of ground state molecules, as relevant to the experiments discussed below (see Fig.~\ref{fig:1}). We consider the behavior of the ground ($J=0$) and first excited rotational state ($J=1$) coupled by a microwave dressing field while exposed to a far-detuned optical field (see Fig.~\ref{fig:1}(a)). The optical field has an  intensity $I$ and $\theta$ is the angle between the polarization vector $\vec{E}_{\mathrm{ODT}}$ and the  magnetic field $\vec{B}$, the latter setting the quantization axis. In our experiments, the magnetic field is sufficiently strong that hyperfine coupling between nuclear spin states is strongly suppressed. Thus, we omit the molecular hyperfine structure. The microwave dressing field has $\mathrm{\sigma^+}$-polarization and couples the states $\ket{0, 0}$ and $\ket{1, 1}$ (see Fig.~\ref{fig:1}(b)). The  couplings between rotational states in this experimental setting are described by the Hamiltonian matrix
\begin{equation*}
\begin{aligned}
& ~~~~~~~ \left | 0,0 \right> ~~~~~~~~~~ \left | 1,0 \right> ~~~~~~~~ \left | 1,-1 \right> ~~~~~~ \left | 1,1 \right> \\
H = 
& \begin{pmatrix}
\hbar\Delta - \alpha_{\mathrm{00}} I & \mathrm{0} & \mathrm{0} & \hbar\Omega/\mathrm{2} \\
\mathrm{0} & \epsilon_{\mathrm{13}} - \alpha_{\mathrm{11}} I & - \alpha_{\mathrm{12}} I & - \alpha_{\mathrm{13}} I \\
\mathrm{0} & - \alpha_{\mathrm{21}} I & \epsilon_{\mathrm{23}} - \alpha_{\mathrm{22}} I & - \alpha_{\mathrm{23}} I \\
\hbar\Omega/\mathrm{2} & - \alpha_{\mathrm{31}} I & - \alpha_{\mathrm{32}} I & - \alpha_{\mathrm{33}} I
\end{pmatrix}
\end{aligned}
\end{equation*}
with~\cite{neyenhuis2012anisotropic}
\begin{align*} 
    \alpha_{\mathrm{11}} &= \alpha_{\mathrm{00}} + \frac{\mathrm{3 cos^2} \theta - \mathrm{1}}{\mathrm{5}}\delta\alpha , \\
    \alpha_{\mathrm{22}} &= \alpha_{\mathrm{33}} = \alpha_{\mathrm{00}} - \frac{\mathrm{3 cos^2} \theta - \mathrm{1}}{\mathrm{10}}\delta\alpha , \\
    \alpha_{\mathrm{23}} &= \alpha_{\mathrm{32}} = - \frac{\mathrm{3 sin^2} \theta}{\mathrm{10}}\delta\alpha ,\, \mathrm{and} \\
    \alpha_{\mathrm{12}} &= \alpha_{\mathrm{21}} = - \alpha_{\mathrm{13}} = - \alpha_{\mathrm{31}} = \frac{\mathrm{3 \sqrt{2} sin} \theta \mathrm{cos} \theta}{\mathrm{10}}\delta\alpha .
\end{align*}
Here, $\epsilon_{\mathrm{13}}$ and $\epsilon_{\mathrm{23}}$ denote the energy difference between $\left | 1,0 \right>$ and $\left | 1,1 \right>$, and $\left | 1,-1 \right>$ and $\left | 1,1 \right>$ at $I = 0$~\cite{SI}, respectively. Light shifts that arise for $I>0$ are characterized by the \emph{isotropic} polarizability $\alpha_{\mathrm{00}}$, leading to a linear light shift for all states, and the \emph{anisotropic} polarizability $\delta\alpha$, leading to a tensor light shift that depends on the polarization angle $\theta$. For NaCs molecules in an optical field at 1064 nm wavelength, we measure $\alpha_{00} = h \times \mathrm{41(1)~kHz/(kW/cm}^2)$~\cite{stevenson2023ultracold} and $\delta \alpha = h \times \mathrm{40(3)~kHz/(kW/cm}^2)$~\cite{SI}. The tensor light shift leads to a mixing of all sublevels in the first excited rotational state, $\left | 1,-1 \right>$, $\left | 1,0 \right>$, and $\left | 1,1 \right>$.

In the absence of a microwave dressing field ($\mathrm{\Omega = 0}$), only the tensor light shift can induce state mixing. In the limit of low optical intensity $I$, off-diagonal mixing is negligible compared to the energy differences of the sublevels. As a result a magic polarization angle appears for $3 \cos^2 \theta_\mathrm{m} = 1$, corresponding to $\theta_\mathrm{m} = 54.7 \degree$, where the three rotational transitions from $J=0$ to $J=1$ become insensitive to variations of the optical intensity. However, in the limit of high optical intensity, off-diagonal state mixing is significant and the magic polarization angle no longer exists. This is the typical scenario for NaCs molecules in optical tweezers~\cite{cairncross2021assembly} or optical dipole traps~\cite{stevenson2023ultracold}, as low intensities are not sufficient to hold NaCs against gravity~\cite{rosenband2018elliptical, SI}.

Surprisingly, in the presence of a sufficiently strong microwave dressing field, state mixing simplifies again and a magic transition can be created. When $\hbar \Omega \gg \delta\alpha I$, which is the case for microwave shielding, the dressing field constitutes the dominant energy scale. State mixing is dominated by the $\sigma^+$ microwave coupling between $|0,0 \rangle$ and $|1,1\rangle$, while admixtures of $|1,-1 \rangle$ and $|1,0 \rangle$ from the tensor light shift are strongly suppressed (see Fig.~\ref{fig:1}(c)). In this limit, the dressed states are given by
\begin{align*}
    \left | \mathrm{+} \right >_n &= \cos(\phi)\left |0,0 ;n \right> + \sin(\phi)\left | 1,1 ;n-1 \right>,\\
    \left | \mathrm{-} \right >_n &= -\sin(\phi)\left | 0,0 ;n \right> + \cos(\phi)\left | 1,1;n-1 \right>,
\end{align*}
and their energy splitting is $\Omega_\mathrm{eff} = \sqrt{\Omega^2 + \Delta^2}$. Here, $\phi$ denotes the mixing angle that is determined via $\sin (2\phi) = 1 / \sqrt{1+(\Delta/\Omega)^2}$ (see Fig.~\ref{fig:1}(d)) and $n$ denotes the photon number manifold of the dressed states \cite{cohen1998atom}, which will be relevant for the discussion of dressed-state spectroscopy below. The polarizability of the dressed states is given by 
\begin{align}
    \label{eq:1}
    \alpha_{\left | \mathrm{+} \right>} &= \mathrm{cos^2}(\phi)\alpha_{00} + \mathrm{sin^2}(\phi)\alpha_{33}, \\
    \alpha_{\left | \mathrm{-} \right>} &= \mathrm{sin^2}(\phi)\alpha_{00} + \mathrm{cos^2}(\phi)\alpha_{33}.
\end{align}
When microwave dressing is used for collisional shielding, the molecules are specifically prepared in the $|+ \rangle$ state. Its polarizability depends on $\phi$, which is controlled via the parameter $\Delta/\Omega$. Here, $\mathrm{\theta_\mathrm{m} = 54.7 \degree}$ becomes a magic polarization angle again, as $\alpha_{00}$ equals $ \alpha_{33}$. The polarizabilities of $| \mathrm{+} \rangle$ and $| \mathrm{-} \rangle $ are identical and, remarkably, transitions between them are magic for any value of $\Delta/\Omega$.

We now experimentally investigate the impact of optical fields on microwave-shielded NaCs ground state molecules. About $2 \times 10^4$ molecules are prepared in a crossed optical dipole trap operating at 1064 nm. The temperature of the molecular gas is 700(100) nK. A homogeneous magnetic field of 864~G points along the $z$-axis. All molecules are prepared in the $|J, m_J \rangle = |0, 0\rangle$ rovibrational ground state and the hyperfine state |$m_{I_\mathrm{Na}}, m_{I_\mathrm{Cs}}\rangle$ = $|3/2 , 3/2 \rangle$, where $m_{I_\mathrm{Na}}$ and $m_{I_\mathrm{Cs}}$ denote the nuclear spin projection of Na and Cs on the quantization axis, respectively. Further details on molecule preparation are discussed in earlier work~\cite{lam2022high, stevenson2023ultracold, warner2023efficient}.
To prepare molecules in the microwave-shielded state $\ket{+}$, a $\sigma^+$-polarized microwave field~\cite{yuan2023planar} is adiabatically increased within $40~\mathrm{\mu s}$ as described in recent work~\cite{bigagli2023collisionally}.

\begin{figure}[]
    \centering
    \includegraphics[width = 8.6 cm]{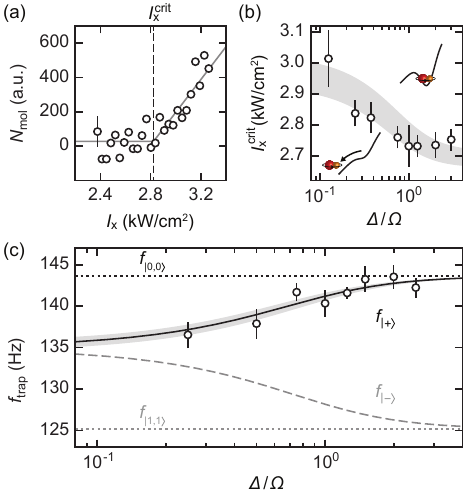}
    \caption{Impact of microwave dressing on optical trapping. Microwave-shielded molecules in the $\left | + \right>$ state are held in a crossed optical dipole trap with a polarization angle of $\theta =0(2)\degree$ for both beams. The Rabi frequency of the microwave field is $\mathrm{\Omega/(2\pi) = 4.0(4)}$ MHz, while the detuning $\Delta$ is varied. (a) Sample measurement of the trap bottom for $\Delta/\Omega = 0.375$. 25 ms after changing the ODTx intensity to $I_\mathrm{x}$, the number of remaining molecules in the trap is detected. The error bar shows a typical error. The critical intensity $I_x^{\mathrm{crit}}$ (black dashed line) below which molecules are untrapped is extracted from a bi-linear fit (gray solid line). (b) Critical intensity $I_x^{\mathrm{crit}}$ as a function of $\Delta/\Omega$. The grey-shaded area shows the prediction based on the trap model. (c) Trap frequency in $z$ direction as a function of $\Delta/\Omega$. The black solid line shows a calculation based on Eq.~(\ref{eq:1}) with the trap frequency of $|0, 0\rangle$ being the only fitting parameter (gray shading shows the uncertainty). The grey dashed line shows the expected behavior for the $\left | - \right>$ state.}
    \label{fig:3}
\end{figure}

First, we study the impact of microwave shielding on optical trapping (see Fig.~\ref{fig:3}). In our setup, molecules in the $| + \rangle$ state are held in the crossed dipole trap with the polarization vector of ODTx and ODTy each pointing along $z$ direction. The trap depth for dressed molecules varies as a function of $\Delta/\Omega$. In the presence of gravity, this leads to a change in the trap bottom, which we observe as shown in Figs.~\ref{fig:3}(a) and (b). The intensity of ODTx, $I_x$, is varied, while keeping the intensity of ODTy constant. Below a critical value, $I_{x}^\mathrm{crit}$, the trap stops supporting the molecules against gravity and they escape from the trap. This change in $I_{x}^\mathrm{crit}$ can have a notable impact on cold molecular clouds. For fixed $I_x = 3.0$~kW/cm$^2$, the trap depth changes from $k_\mathrm{B} \times 100$~nK to zero while $\Delta/\Omega$ is varied from $3$ to $0.15$. This is a drastic effect for BECs of microwave-shielded molecules at a temperature of about 10~nK \cite{bigagli2024observation} and becomes particularly important when the parameter $\Delta/\Omega$ is used to tune the effective strength of dipole-dipole interactions between dressed molecules \cite{schindewolf2022evaporation, bigagli2023collisionally, lin2023microwave}. 

The change in the trapping potential for shielded molecules is also reflected in the trap frequency (see Fig.~\ref{fig:3}(c)). Via parametric loss resonance, we measure the trap frequency in the $z$-direction as a function of $\Delta/\Omega$~\cite{SI}. For our setup, we observe a reduction in trap frequency by about 6$\%$ when $\Delta/\Omega$ changes from 2.5 to 0.25, which arises from the change of polarizability $\alpha_{\left | + \right>}$. Employing an accurate model of the optical dipole trap and taking into account Eq.~(\ref{eq:1}), we find excellent agreement with the measured trap frequencies.

\begin{figure}[]
    \centering
    \includegraphics[width = 8.6 cm]{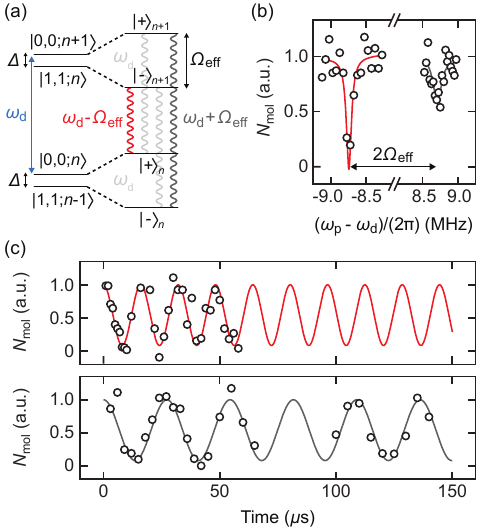}
    \caption{Mollow triplet transitions between microwave-dressed rotational states in NaCs. (a) Illustration of the allowed transitions between the dressed states. The left part shows the states in the uncoupled basis of molecular states and microwave photon numbers. The right part shows states in the microwave-dressed basis. Wavy lines show the allowed transitions which can be driven by an additional microwave probe field. Red and grey wavy lines denote the two sideband transitions ($\mathrm{\omega_p =\omega_d \pm \Omega_{eff}}$); light grey wavy lines denote the center transitions ($\mathrm{\omega_p = \omega_d}$). (b) Out-of-trap spectroscopy on the microwave-shielded state $\left | + \right>_n$ at $\Delta/\Omega = 0.24$ and $\Omega / (2 \pi) = 8.49(3)$ MHz, illuminated by the probe microwave field for 5 $\mathrm{\mu s}$. The corresponding mixing angle is $\phi = 38\degree$, ensuring the transition moments are strong. (c) Coherent Rabi oscillations between dressed rotational states. The upper (lower) panel corresponds to the $\mathrm{\omega_d-\Omega_{eff}}$ ($\mathrm{\omega_d+\Omega_{eff}}$) sideband. The sinusoidal fits give Rabi frequencies of 59.6(18)~kHz and 35.2(9)~kHz, respectively.} 
    \label{fig:5}
\end{figure}

Next, we introduce dressed-state spectroscopy for microwave-dressed molecules. This technique allows us to precisely measure energy shifts between dressed states. Utilizing a separate microwave probe field, we can drive transitions between the $|+\rangle$ and the $|-\rangle$ state. As $\Omega$ is much larger than the linewidth of the dressed states, spectral side bands arise, as illustrated in Fig.~\ref{fig:5}(a). The transition dipole moment is zero between dressed states with the same photon number $n$ and nonzero between dressed states whose photon number differs by 1. This leads to four allowed transitions with three distinct resonance frequencies, known as the Mollow triplet \cite{cohen1998atom}. These transitions can only be driven by a $\sigma^+$-polarized microwave field and the strength of the transitions is proportional to the transition dipole moments that depend on the mixing angle $\phi$ as: 
\begin{gather}
    \label{eq:3}
    \prescript{}{n+1}{\left < -|\hat{d}| + \right >_n} = \mathrm{cos(\phi)cos(\phi)} d_\mathrm{0}/\sqrt{3} , \\
    \label{eq:4}
    \prescript{}{n-1}{\left < -|\hat{d}| + \right >_n} = - \mathrm{sin(\phi)sin(\phi)} d_\mathrm{0}/\sqrt{3} , \\
    \prescript{}{n+1}{\left < +|\hat{d}| + \right >_n} = 
    - \prescript{}{n+1}{\left < -|\hat{d}| - \right >_n} = \mathrm{sin(\phi)cos(\phi)} d_\mathrm{0}/\sqrt{3}.
\end{gather}
Here, $\hat{d}$ is the dipole operator, and $d_{\mathrm{0}} = 4.6(2)$ D~\cite{dagdigian1972molecular} is the electric dipole moment of NaCs in the molecule-fixed frame. In the absence of an optical field, the transition frequencies are $\mathrm{\omega_d-\Omega_{eff}}$, $\mathrm{\omega_d+\Omega_{eff}}$, and $\mathrm{\omega_d}$, respectively, where $\mathrm{\omega_d}$ denotes the frequency of the microwave dressing field. 

\begin{figure}[]
    \centering
    \includegraphics[width = 8.6 cm]{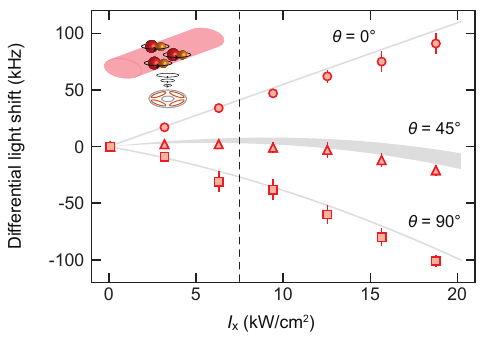}
    \caption{Differential light shifts of the lower sideband transition ($\mathrm{\omega_d-\Omega_{eff}}$) between dressed states for microwave-shielded molecules at $\Delta/\Omega = 0.95$ and $\Omega / (2 \pi) = 5.28(1)$ MHz. Positive shifts denote larger transition frequencies. Circle, triangle, and square markers stand for $0\degree$, $45\degree$, and $90\degree$ optical polarization angles, respectively. The light grey area represents the calculation with a $2 \degree$ uncertainty of the polarization angle. The dashed black line indicates the presence of a magic intensity around $\mathrm{7.5(5)~kW/cm^2}$ for $\theta = 45 \degree$.} 
    \label{fig:6}
\end{figure}

We probe the transitions of the Mollow triplet of microwave-shielded NaCs, first in the absence of an optical field. By varying the frequency of the microwave probe field $\omega_\mathrm{p}$ and monitoring the number of molecules that remain in the $\left | + \right>$ state, we observe the two sideband transitions with frequencies $\mathrm{\omega_d-\Omega_{eff}}$ and $\mathrm{\omega_d+\Omega_{eff}}$. A typical spectrum is shown in Fig.~\ref{fig:5}(b). The different depth of the resonance dips is reflecting the different coupling strengths of the sideband transitions, which we analyze in detail below. The two center transitions at frequency $\omega_\mathrm{d}$ are not observed, as our detection method is only sensitive to the molecular state, which does not change upon driving these transitions. Dressed-state spectroscopy provides a highly precise method for calibrating the coupling strength $\Omega$ of the microwave dressing field which we made extensive use of to characterize the microwave dressing fields in Ref.~\cite{bigagli2024observation}. The microwave probe also allows coherent driving of transitions between the dressed states $\left | + \right>$ and $\left | - \right>$. As shown in Fig.~\ref{fig:5}(c), we observe coherent Rabi oscillations. By extracting the frequency of Rabi oscillations of the sideband transitions $\mathrm{\omega_d \pm \Omega_{eff}}$, we  measure the relative strength of the Mollow triplet transitions. For the data in Fig.~\ref{fig:5}(c), the ratio of Rabi frequencies is 1.7(1), which agrees with the theoretical expectation of 1.6 for $\Delta/\Omega = 0.24$ from Eqs.~(\ref{eq:3}) and (\ref{eq:4}). 

Finally, we utilize dressed-state spectroscopy to measure the differential light shift between $\left | + \right>$ and $\left | - \right>$ in the presence of the optical trapping field. As discussed above, we expect the emergence of a magic transition and we demonstrate this effect here experimentally. The shielding parameter is set to $\Delta/\Omega = 0.95$, where microwave shielding with a $\sigma^+$ field works most effectively and a lifetime of the molecular sample of about 1 s can be reached \cite{bigagli2023collisionally}. We probe the lower sideband transition $\mathrm{\omega_d-\Omega_{eff}}$ in the presence of ODTx and keep track of its resonance frequency for various intensities $I_x$ and polarization angles $\theta$. This measures the differential light shift of this transition as shown in Fig.~\ref{fig:6}. In contrast to molecules without a dressing field, molecules under microwave shielding show a magic angle of $\theta = 45(2) \degree$ around $I_x = \mathrm{7.5(5)~kW/cm^2}$. This angle deviates from the theoretical prediction $\theta_\mathrm{m} = 54.7 \degree$ (see above), as the dressing field is not sufficiently strong compared with the optical field and higher-order perturbations need to be taken into account. We discuss additional details in~\cite{SI}. Based on the measured differential light shifts, we expect that a coherence time around 400 ms is possible, similar to the recent demonstrations of magic ellipticity~\cite{park2023extended} and magic wavelength~\cite{gregory2024second}. This approach may prove useful for the realization of rotational qubits or long-lived effective spin systems of dipolar molecules. We note that a similar microwave dressing technique has recently been demonstrated to extend the coherence time of spin qubits in silicon \cite{laucht2017dressed, miao2020universal}. 

In conclusion, we have discussed the interplay of microwave dressing, optical polarization, and optical intensity and their impact on the single particle properties of optically trapped microwave-shielded molecules. Microwave dressing gives rise to a tunable ladder of dressed rotational states that enables precision spectroscopy and magic rotational transitions. Conceptually, this work illustrates that microwave-dressed molecules can be viewed as a new object, jointly comprised of a molecule and a microwave photon field. Compared to bare ground state molecules, dressed molecules have highly modified single particle and collisional properties. While these characteristics are similar to polaritons, it is worth noting that microwave-dressed molecules are long-lived on the single molecule level. 

Looking forward, dressed-state spectroscopy and magic transitions promise to become powerful tools for probing microwave-shielded molecules. Dressed-state spectroscopy has already been used in Ref.~\cite{bigagli2024observation} to measure the Rabi couplings and relative orientation of $\sigma^+$- and $\pi$-polarized microwave fields to realize double microwave shielding. In the future, we expect dressed-state spectroscopy of molecular systems to play a similar role as radio-frequency spectroscopy in atomic systems, for example, enabling precision studies of interaction shifts in bulk and lattice systems of dipolar molecules~\cite{hazzard2011spectroscopy, baier2016extended, tobias2022reactions, li2023tunable}. Magic dressed-state transitions also suggest new ways to realize highly coherent rotational qubits and long-lived rotational spin systems, e.g.~in optical lattices~\cite{gorshkov2011tunable, yan2013observation, christakis2023probing, lin2022seconds} and tweezers~\cite{holland2023demand, bao2023dipolar, ruttley2024enhanced,  picard2024site}. 

We thank Silvia Cardenas-Lopez and Ricardo Gutierrez-Jauregui for helpful discussions, and Emily Bellingham and Tarik Yefsah for critical reading of the manuscript. This work was supported by an NSF CAREER Award (Award No.~1848466), an ONR DURIP Award (Award No.~N00014-21-1-2721), and a Lenfest Junior Faculty Development Grant from Columbia University. C.W. acknowledges support from the Natural Sciences and Engineering Research Council of Canada (NSERC). W.Y.\ acknowledges support from the Croucher Foundation. I.S. was supported by the Ernest Kempton Adams Fund. S.W.\ acknowledges additional support from the Alfred P. Sloan Foundation.


%


\end{document}


\title{Supplemental Material for ``Dressed-State Spectroscopy and Magic Trapping of Microwave-Shielded NaCs Molecules''}

\preprint{APS/123-QED}

\author{Siwei Zhang}
\affiliation{Department of Physics, Columbia University, New York, New York 10027, USA}
\author{Weijun Yuan}
\affiliation{Department of Physics, Columbia University, New York, New York 10027, USA}
\author{Niccol\`{o} Bigagli}
\affiliation{Department of Physics, Columbia University, New York, New York 10027, USA}
\author{Claire Warner}
\affiliation{Department of Physics, Columbia University, New York, New York 10027, USA}
\author{Ian Stevenson}
\affiliation{Department of Physics, Columbia University, New York, New York 10027, USA}
\author{Sebastian Will}\email{Corresponding author. E-mail: sebastian.will@columbia.edu}
\affiliation{Department of Physics, Columbia University, New York, New York 10027, USA}
\date{\today}

\maketitle

\section{S1. DIPOLE TRAP DESCRIPTION}
The optical trapping light is generated by a 1064 nm narrow-line single-mode Nd:YAG laser (Coherent Mephisto MOPA). The trap has been precisely characterized using Cs atoms, which, together with information on the beam parameters, allows us to build a precise numerical model of the trap~\cite{warner2021overlapping}. The model is adapted to NaCs molecules via the known relative polarizability between Cs atoms and NaCs molecules~\cite{stevenson2023ultracold}. For the data in Fig.~2 in the main text, the $x$-dipole trap (ODTx) is elliptical and focused to waists of 149(6)~$\mu$m (horizontal) and 52(6)~$\mu$m (vertical); the $y$-dipole trap (ODTy) is circular with a waist of 165(5)~$\mu$m. For the rest of the data, the ODTx is still elliptical but focused to waists of 127(5)~$\mu$m (horizontal) and 56(3)~$\mu$m (vertical); ODTy is circular with a waist of 106(5)~$\mu$m. 

\section{S2. PARAMETRIC LOSS RESONANCE}
\begin{figure}[H]
    \centering
    \includegraphics[width = 8.6 cm]{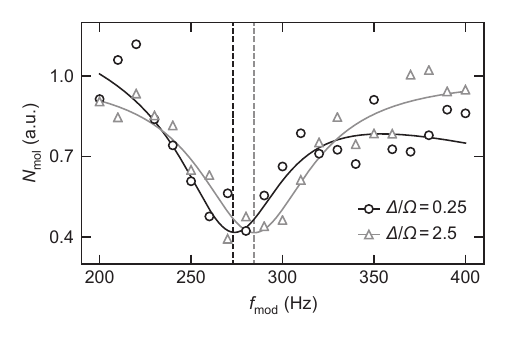}
    \caption{Exemplary raw data for Fig.~2(c) in the main text, measuring trap frequencies via parametric loss resonance. Black (grey) circles (triangles) show data for $\Delta/\Omega = 0.25$ (2.5). The solid lines show Lorentzian fits. Dashed vertical lines mark the resonance positions from the fit.}
    \label{fig:s5}
\end{figure}

We use the parametric loss resonance method to measure the trap frequency of NaCs in the crossed dipole trap. We synchronously modulate the intensities of ODTx and ODTy by {$\pm 30$\%} with a frequency $f_{\mathrm{mod}}$ for 100 ms. When $f_{\mathrm{mod}} = 2 f_\mathrm{trap}$, molecules are parametrically excited. Resonant excitation is detected as loss from the trap, as shown in Fig.~\ref{fig:s5}.

\section{S3. DETERMINATION OF ANISOTROPIC POLARIZABILITY}

\begin{figure}[]
    \centering
    \includegraphics[width = 8.6 cm]{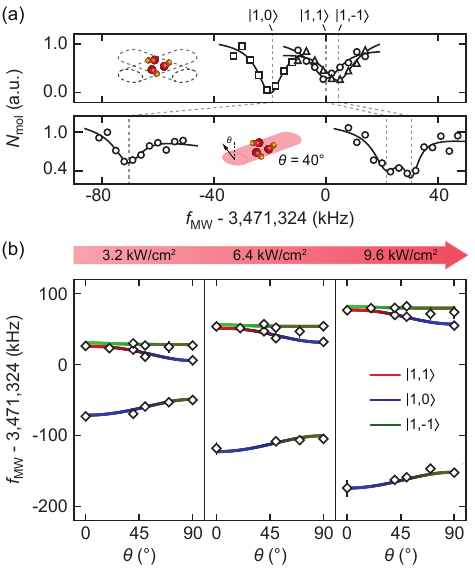}
    \caption{Measurement of the anisotropic polarizability of NaCs molecules in the \emph{absence of microwave dressing}. (a) Microwave absorption spectra without (upper panel) and with (lower panel) an optical field present. Dashed vertical lines mark the location of observed resonances. Square, circle, and triangle markers represent the polarization of the probe microwave antenna set to $\mathrm{\pi}$, $\mathrm{\sigma^+}$, and $\mathrm{\sigma^-}$, respectively. Data in the lower panel is taken at optical intensity $I_x$ = 3.2~kW/cm$^2$ and polarization angle $\theta = 40 \degree$ (optical intensities are calibrated using Cs atoms). (b)  Differential light shift of rotational transitions versus optical polarization angles. Polarization angles have an uncertainty of $\pm 2\degree$. Each data point corresponds to a measured resonant transition frequency, showing the relative light shifts between $J$ = 0 and the $J$ = 1 sublevels. The three panels show different optical intensities of the dipole trap. The color gradients of the fitting curves encode the admixing of the different $J$ = 1 rotational sublevels (see legend).}
    \label{fig:S8}
\end{figure}

Here, we discuss our measurements of isotropic and anisotropic optical polarizability of NaCs molecules. These measurements are performed in the \emph{absence of microwave dressing} with an optical field at 1064 nm. 

The isotropic polarizability is $\alpha_{00} = h \times \mathrm{41(1)~kHz/(kW/cm}^2)$, as we measured in earlier work~(Ref.~\cite{stevenson2023ultracold}). To measure the anisotropic polarizability $\delta\alpha$ we perform microwave spectroscopy that is sensitive to differential light shifts between $J=0$ and $J=1$. After preparation in the crossed dipole trap, the molecules are transferred to a single beam ODT along the $x$-direction with variable polarization angle. A weak microwave pulse is applied for 150 $\mathrm{\mu s}$ and the population of molecules in $\left | 0,0 \right>$ is monitored as a function of microwave frequency. The resulting spectra reveal the energy spacing between $\left | 0,0 \right>$ and the three $m_J= -1, 0, +1$ sublevels in $J=1$. Fig.~\ref{fig:S8}(a) shows examples of microwave spectra without ODTx (upper panel) and with ODTx (lower panel). 

From the spectrum without the optical field, we fit the rotational energy splittings to be $\epsilon_{13} = - h \times 20(2)$~kHz and $\epsilon_{23} = h \times 5(1)$~kHz, for the energy difference between $\left | 1,0 \right>$ and $\left | 1,1 \right>$, and $\left | 1,-1 \right>$ and $\left | 1,1 \right>$, respectively. 

For the spectra with the optical field, the anisotropic polarizability leads to a mixing of the excited $m_J$ sublevels, which results in differential energy shifts. We have characterized these shifts for various polarization angles and ODTx intensities by measuring the respective resonance frequencies. The results are shown in Fig.~\ref{fig:S8}(b). Using the anisotropic polarizability as the only free fitting parameter, we obtain a best-fit value of $\delta \alpha = h \times \mathrm{40(3)~kHz/(kW/cm}^2)$. 

Using the definition of isotropic and anisotropic polarizabilities,
$\alpha_{00} = \frac{1}{3}(\alpha_{\parallel} + 2\alpha_{\perp})$ and $\delta\alpha = \frac{2}{3}(\alpha_{\parallel} - \alpha_{\perp})$, we extract the polarizability along the internuclear axis, $\alpha_{\parallel} = h \times \mathrm{81(6)~kHz/(kW/cm}^2)$, and the polarizability perpendicular to it, $\alpha_{\perp} = h \times \mathrm{21.4(13)~kHz/(kW/cm}^2)$. Specifically, for a polarization angle of $\theta =0$, the polarizabilities in states $|1,0 \rangle$, $|1,-1 \rangle$, and $|1,1 \rangle$ are measured as $\alpha_{11} = h \times \mathrm{57(3)~kHz/(kW/cm}^2)$, $\alpha_{22} = h \times \mathrm{33(1)~kHz/(kW/cm}^2)$, and $\alpha_{33} = h \times \mathrm{33(1)~kHz/(kW/cm}^2)$, respectively. The  ratio of polarizabilities of $|1,1 \rangle$ and $|0,0\rangle$ is $\gamma = \alpha_{33}/\alpha_{00} = 0.807(12)$.

\section{S4. CALCULATION OF MAGIC CONDITIONS FOR BARE {NaCs} MOLECULES}

\begin{figure}[]
    \centering
    \includegraphics[width = 8.6 cm]{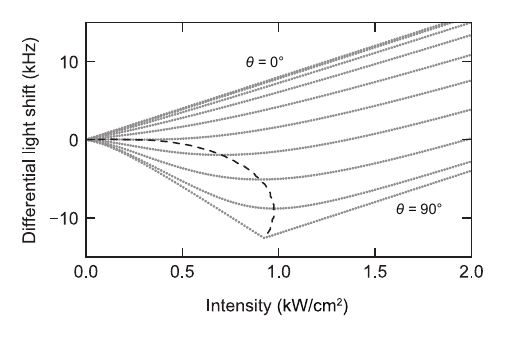}
    \caption{Magic conditions for NaCs molecules exposed to an optical field at 1064 nm, but without microwave dressing. Grey dotted lines show the differential light shift for the rotational transition between $|0, 0\rangle$ and $|1, 1\rangle$ as a function of optical intensity for different polarization angles from $0\degree$ to $90\degree$ (top to bottom), spaced by $10\degree$. The black dashed curve indicates the positions of the magic condition for each polarization angle, where the differential light shift is first-order insensitive to fluctuations of the optical field. For all polarization angles, the intensities for magic trapping are smaller than $\mathrm{1.0(1) ~ kW/cm^2}$.}
    \label{fig:s6}
\end{figure}

Taking into account the known polarizabilities for NaCs, we calculate the conditions under which magic trapping can be achieved in the \emph{absence of a microwave dressing field}. To this end, we diagonalize the coupling matrix shown in the main text and obtain the differential light shift of a rotational transition for various linear polarization angles of the optical field. Fig.~\ref{fig:s6} shows this calculation for the rotational transition between $|0, 0\rangle$ and $|1, 1\rangle$. We find that magic conditions can only be achieved for optical intensities below $\mathrm{1.0(1) ~ kW/cm^2}$. For NaCs molecules in optical tweezers or optical dipole traps at 1064 nm, such intensities are not sufficient to support NaCs against gravity in typical experiments. 

\section{S5. CALCULATION OF THE DIFFERENTIAL LIGHT SHIFT FOR DIFFERENT MICROWAVE DRESSING FIELDS}

\begin{figure}[]
    \centering
    \includegraphics[width = 8.6 cm]{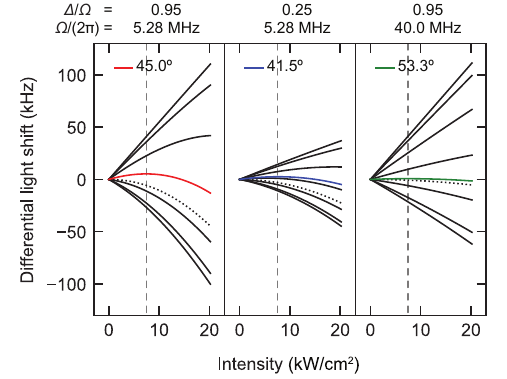}
    \caption{Effect of microwave dressing on the differential light shift and magic angle, for the lower sideband transition ($\mathrm{\omega_d-\Omega_{eff}}$). From left to right, calculations for three different microwave parameters are shown, as indicated above the plot. For each panel, the colored curve shows the shifts for the polarization angle that is magic at the intensity of 7.5 kW/cm$^2$, the location of which is marked by the grey dashed line. For comparison, the black dotted lines show the theoretical magic angle of $\theta_{\mathrm{m}} = 54.7\degree$ in the limit of strong microwave dressing. The black solid lines show the results for polarization angles from $0\degree$ to $90\degree$, spaced by $15\degree$ (from top to bottom).}
    \label{fig:s3}
\end{figure}

In Fig.~4 of the main text, we show the measurement of differential light shifts for NaCs in the \emph{presence of a microwave dressing field} with $\Omega/(2 \pi) = 5.28$ MHz and $\Delta/\Omega = 0.95$. We find that the magic angle at an optical intensity of $I_{x} = 7.5$ kW/cm$^2$ is at $\theta = 45(2)^\circ$, which deviates from the theoretical prediction  $\theta_\mathrm{m} = 54.7^\circ$ in the limit of extremely strong microwave dressing. This difference arises as the microwave dressing energy is only about ten times stronger than the optical tensor shift, such that the tensor light shift still affects the differential light shift around the magic angle. 

To provide additional details on this point, Fig.~\ref{fig:s3} shows calculations of differential light shifts at different polarization angles for different microwave detunings $\Delta$ and different Rabi frequencies $\Omega$. The closer the microwave is to resonance (smaller $\Delta$), the smaller the differential shifts are, as the rotational state composition of the two dressed states becomes more similar. The stronger the microwave field, the higher the suppression of higher-order effects, and the closer the magic angle is to the theoretical value $\theta_{\mathrm{m}} = 54.7\degree$.

\section{S6. ESTIMATION OF THE COHERENCE TIME}

\begin{figure}
    \centering
    \includegraphics[width = 8.6 cm]{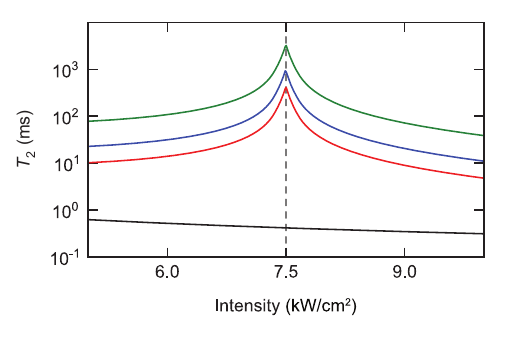}
    \caption{Projected coherence times $T_{2}$ at three different microwave settings and the corresponding magic polarization angles at an optical intensity of 7.5~kW/cm$^2$ (as in Fig.~\ref{fig:s3}). From top to bottom, data is shown for $\Omega/(\mathrm{2\pi}) = 40.0$ and $\Delta/\Omega = 0.95$ (green); $\Omega/(\mathrm{2\pi}) = 5.28$ MHz and $\Delta/\Omega = 0.25$ (blue); $\Omega/(\mathrm{2\pi}) = 5.28$ MHz and $\Delta/\Omega = 0.95$ (red); and bare molecules without microwave dressing at optical polarization angle $\theta = 0 \degree$ (black). The peak coherence times at 7.5~kW/cm$^2$ are 3300 ms, 960 ms, 420 ms, and 0.42 ms, respectively.}
    \label{fig:s4}
\end{figure}

Based on the measured differential light shifts, we can make a projection for the coherence time that can be achieved for superposition states of $\left | + \right>_n$ and $\left | - \right>_{n+1}$ under magic conditions. Assuming $\Delta/\Omega \sim 1$, a trap intensity $I_x = 7.5$~kW/cm$^2$, and the magic angle $\theta = 45 \degree$, we calculate a transition frequency spread of $\Delta f \sim 2.4~\mathrm{Hz}$ when the molecular gas samples a $4\%$ variation of optical intensity across the cloud size (which is a typical value for a dipole trap, e.g.~also see~\cite{blackmore2020controlling}). For bare molecules at $\theta = 0$, the transition frequency spread is $\Delta f_{\rm{b}} \sim 2400~\mathrm{Hz}$ and the corresponding coherence time is limited to $1 / \Delta f_{\rm{b}} \sim 0.4~$ms. With the reduced frequency spread under magic conditions, we expect a coherence time around 400 ms, similar to the recent demonstrations of magic ellipticity \cite{park2023extended} and magic wavelength~\cite{gregory2024second}. 

Compared to the estimate above, it is possible to further extend the coherence time by using stronger Rabi frequencies or smaller detunings $\Delta$. Fig.~\ref{fig:s4} shows estimates for the coherence time $T_{2}$ at the magic angle for the three microwave parameters of Fig.~\ref{fig:s3}. The coherence time becomes longer for dressing fields with higher Rabi coupling and closer proximity to resonance. Our projection indicates that the coherence time can be increased by three to four orders of magnitude compared to bare molecules, especially by utilizing strong dressing fields. The practically achievable coherence time is limited by amplitude fluctuations of the microwave dressing field. The transition frequency between dressed states fluctuates with the Rabi frequency of the microwave dressing field, $\Omega$ (see Fig.~3(a) of the main text). This effect can be mitigated by employing active power stabilization or dynamical decoupling sequences~\cite{gullion1990new} in the future.


%